# THE DEVELOPMENT OF A PROBABILISTIC MODEL FOR THOLIN AGGREGATION IN TITAN'S ATMOSPHERE.

C. C. Harris[1,2], L. S. Matthews[1], T. W. Hyde[1], [1]Center for Astrophysics, Space Physics and Engineering Research (CASPER), Baylor University, One Bear Place 97310, Waco, Texas 76798-7310, USA (Lorin_Matthews@baylor.edu, Truell_Hyde@baylor.edu), [2]Randolph College, 2500 Rivermont Ave., Lynchburg, VA, 24503, USA (ccharris@randolphcollege.edu).

**Introduction:** Titan is one of the more distinctive bodies in our solar system. In addition to being the largest of Saturn's moons, its thick atmosphere generates interest because of its similarities and differences with Earth [1, 2]. Like Earth, Titan's lower atmosphere contains clouds which precipitate as rain [2]. This rain forms lakes and rivers of liquid methane and ethane which erode and shape the surface, much like water does on Earth, before evaporating into the atmosphere [2]. In Titan's atmospheric system, a single dominating factor controls the weather, the concentration of the organic molecule tholin.

High concentrations of tholin in the lower atmosphere give Titan its characteristic hazy reddish-orange appearance. As tholin settles out of the formation zones in the upper atmosphere, it condenses into a thick layer at about 80 kilometers above the surface where it dominates the atmospheric activity by filtering the incoming ultraviolet rays (thus regulating temperature) and controlling the atmospheric circulation. Since tholin is also highly influential in the amount of methane present in the atmosphere, methane clouds are influenced by the distribution of tholin concentrations.

A broad and diverse interest in tholin production (from atmospheric dynamics, chemistry, to climatology) spurred the study of the organic aerosol beginning in the late 1970's [3]. Since then, the molecule has been the subject of much research [4], culminating January 14, 2005, when the Huygens probe associated with the Cassini spacecraft descended 1500 km through Titan's atmosphere to the surface. The probe collected valuable information which is now being used to strengthen and improve tholin associated models [5, 6].

Due to the continous influx of galactic cosmic rays and solar UV rays, the particles in Titan's atmosphere are charged and can be modeled as a plasma environment. Huygens measured ion densities and temperature during its mission, important parameters for recreating the plasma conditions in different sections of the atmosphere [6].

The charging mechnismas differ on the day- and night sides of Titan. Photoionization by solar UV rays is the primary charging mechanism during the day, while electron/ion collisions with the ambient plasma dominate nighttime charging [7]. The charging of the grains is further complicated by the fact that the particles are irregular, fractal aggregates. The arrangement of charge on the irregular surface affects the orientation of colliding aggregates, changing the coagulation rate and the resultant morphology of the grains (see Fig. 1) [8].

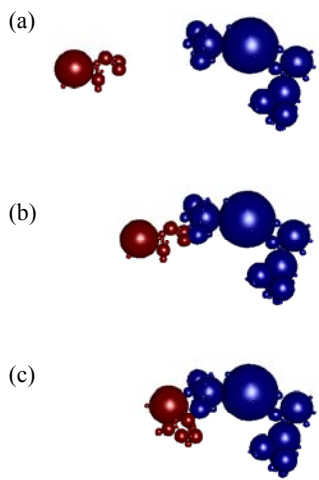

Fig. 1 Result of collisions including dipole-dipole interactions. (a) Initial orientation of two aggregates. (b) Resulting collision when the two aggregates interact electrostatically but do not exert torques on each other. (c) Final orientation when electrostatic forces include the dipole-dipole interactions. The aggregates exert torques on each other and the resultant aggregate is more compact.

Most coagualation simulation models use a statistical approach to calculate the growth of particles at various points in the atmosphere. Such analytical approaches can provide growth rates and particle sizes; however, in these models collisions between particles must occur according to a specific set of rules designed to simplify the physics involved. Usually the particles are approximated as spheres when calculating charge and coagulation probabilties. This approximation is contained in the coagulation kernel in the Smoluchowski equation, which gives the coagulation probability for particles with distinct masses [4]. As a result, self-consistent N-body simulations are essential for properly modeling the dynamics of such interacting grains while also resolving collisions. This study is concerned with modeling the micro-physical processes which govern the aggregation, in particular the charge-dipole interactions. Statistical results from this study will be used to refine the coagulation kernel used in subsequent models of Titan's atmosphere.

**Method:** Tholin is a polycyclic aromatic hydrocarbon formed by photolysis in Titan's upper atmosphere and grows through coagulation of the nucleated embroys. Following the model outlined by Bar-Nun [9], tholin formation is considered in this work with respect to altitude and aggregate mass. As such, the formation of tholin molecules is presented in three stages based on the number of components contained in the aggregate and its associated altitude.

Aggregates were created by numerically modeling interactions between colliding particle pairs [8, 10]. Work was done in the center of mass (COM) frame of an initial seed particle; a second monomer or aggregate was chosen to approach the COM plus an offset from a random direction. The relative velocities between the grains was then determined assuming Brownian motion.

Aggregates of increasing mass were built in several steps by specifying plasma parameters, velocities, and general particle characteristics specific to given altitudes in Titan's atmosphere. The force acting on particle $i$ from the electric field of particle $j$ is calculated along with the torque induced by the dipole-dipole interactions. The torque induces a rotation of the aggregate altering its orientation during collision which can affect the resulting structure.

Collisions are detected when monomers within each aggregate physically overlap. Colliding aggregates are assumed to stick at their point of contact with orientation being preserved. Fragmentation during collisions is not considered due to the low velocities imposed on incoming grains [11]. The charge and dipole moment of resultant aggregates are determined by the use of a heuristic charging scheme [10].

The initial population of embryos were modeled as spheres with radius $r = 40 – 50$ nm. The charges on the particles were calculated for two cases: day-time charging, where UV photoionization is important, and night-time charging, where plasma (formed primarily by ionizing cosmic rays) is collected by the grains. The ion species in the plasma were mass-averaged and adjusted by the percent abundance in calculating the positive charging currents to a grain.

Aggregate populations were grown in three stages. First generation aggregates were created by addition of single monomers until the number of monomers reached $N = 50$. This growth is considered to take place at 270 km altitude and $T = 175$ K. As these particles settle lower in the atmosphere, the second generation forms at an altitude of 200 km and $T = 158$ K, and is grown to size $N = 200$ monomers. The third generation is grown at an altitude of 100 km and $T = 140$ K up to to a size $N = 2500$ monomers.

Collision data collected from aggregate formation include the number of monomers, radius, charge, relative velocity, and fractal dimension for each aggregate. Ultimately this data will be used to determine the coagulation kernel which gives the probability of coagulation for particles with masses $m$ and $m'$.

**Results:** Sample data for the formation of first generation aggregates for each of the charging classes are shown in Figure 2. Initial results indicate that UV photoionization produces more highly-charged aggregates, reducing the collision probability between grains.

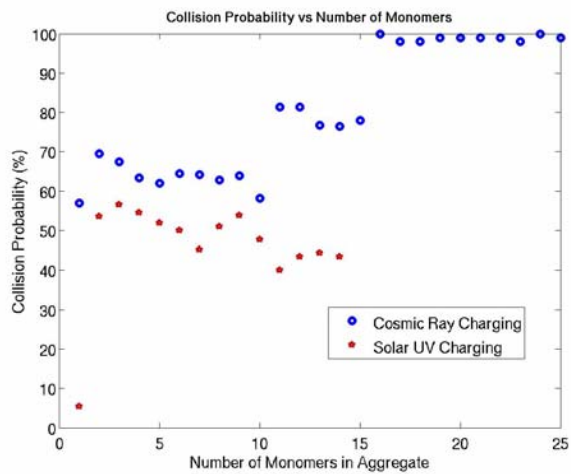

Fig. 2 Collision probability for the first generation of aggregates with respect to the number of monomers contained in the aggregate.

**Conclusion:** Further research is needed to understand the role tholin plays in Titan's tropospheric processes. This study refines the simulation of the growth of tholin particles by including dipole-dipole interactions of the charged aggregates and determining the manner in which they affect coagulation rates. Statistical data including the radius, charge, and collision probability for aggregates with a given mass will eventually be used to calculate the coagulation kernel in the Smoluchowsky equation.